\documentclass[aps, prl,twocolumn,a4paper, superscriptaddress]{revtex4}
\usepackage{amssymb}
\usepackage{amsmath}
\usepackage{epsfig}
\usepackage{graphicx}
\usepackage{dcolumn}
\usepackage{bm}

\begin{document}

\title{Spin-Orbit-Mediated Spin Relaxation in Graphene}
\author{D. Huertas-Hernando}
\affiliation{Department of Physics, Norwegian University of Science and Technology,
NO-7491, Trondheim, Norway}
\author{F. Guinea}
\affiliation{Instituto de Ciencia de Materiales de Madrid, CSIC, Cantoblanco E28049
Madrid, Spain}
\author{Arne Brataas}
\affiliation{Department of Physics, Norwegian University of Science and Technology,
NO-7491, Trondheim, Norway}

\begin{abstract}
We investigate how spins relax in intrinsic graphene. The spin-orbit
coupling  arises from the band structure and is enhanced by ripples.
The orbital motion is influenced by scattering centers and
ripple-induced gauge fields. Spin relaxation due to Elliot-Yafet and
Dyakonov-Perel mechanisms and gauge fields in combination with
spin-orbit coupling are discussed. In intrinsic graphene, the
Dyakonov-Perel mechanism and spin flip due to gauge fields dominate and the
spin-flip relaxation time is inversely proportional to the elastic
scattering time. The spin-relaxation anisotropy depends on an
intricate competition between these mechanisms. Experimental
consequences are discussed.
\end{abstract}

\maketitle


Graphene can be useful in future advanced applications because of the
reduced dimensionality, the long mean free paths and phase coherence
lengths, and the control of the number of carriers \cite{NGPNG08}. Among
possible applications, graphene is investigated as a material for spintronic
devices \cite{HGNSB06,HJPJW07,Hetal07,CCF07,WPLCWSK08,JPTJW08,Ilani08}.
Spintronics aims to inject, detect, and manipulate the electron spin in
electronic devices.

Spin manipulation via the spin-orbit (SO) coupling has been extensively
discussed in semiconductors and metals \cite{ZFS04}. The SO coupling
enables electric, and not just magnetic, control of the spin \cite{KM05}. In two dimensional (2D) semiconducting structures,
inversion asymmetry results in the Rashba SO coupling \cite{Rashba60}%
. Additionally, bulk inversion asymmetry in A$_{3}$B$_{5}$ compounds causes
the Dresselhaus SO coupling \cite{Dresselhaus55}. Device performance
is limited by spin relaxation and understanding its origin enables enhanced
spin control. Two mechanisms of spin relaxation discussed in the literature
\cite{ZFS04,Fabian99}, the Elliof-Yafet \cite{Elliot54,Yafet63} and Dyakonov-Perel \cite{Dyakonov71,Dyakonov86} mechanisms, can be relevant in graphene.

Elliof-Yafet (EY) spin relaxation is related to how the spin changes its
direction during a scattering event \cite{Elliot54,Yafet63}. This is
possible because the SO coupling produces electronic wave functions
that are admixtures of spin and orbital angular momentum. Dyakonov-Perel (DP) \cite{Dyakonov71,Dyakonov86} spin
relaxation is related to spin precession between scattering events by the
effective (Zeeman) magnetic field induced by the SO coupling. This
SO induced effective (Zeeman) magnetic field changes direction
during scattering. In the EY mechanism, the spin
relaxation time is proportional to the elastic scattering time $\tau_\text{el},
\tau^{EY}_\text{so} \propto \tau_\text{el}$, whereas  the dependence is
opposite $\tau^{DP}_\text{so}\propto (\tau_\text{el})^{-1}$ for the DP mechanism. This qualitative difference
allows detection of these two competing mechanisms in disordered samples.

\begin{figure}[!t]
\begin{center}
\includegraphics[width=8.0cm,angle=0]{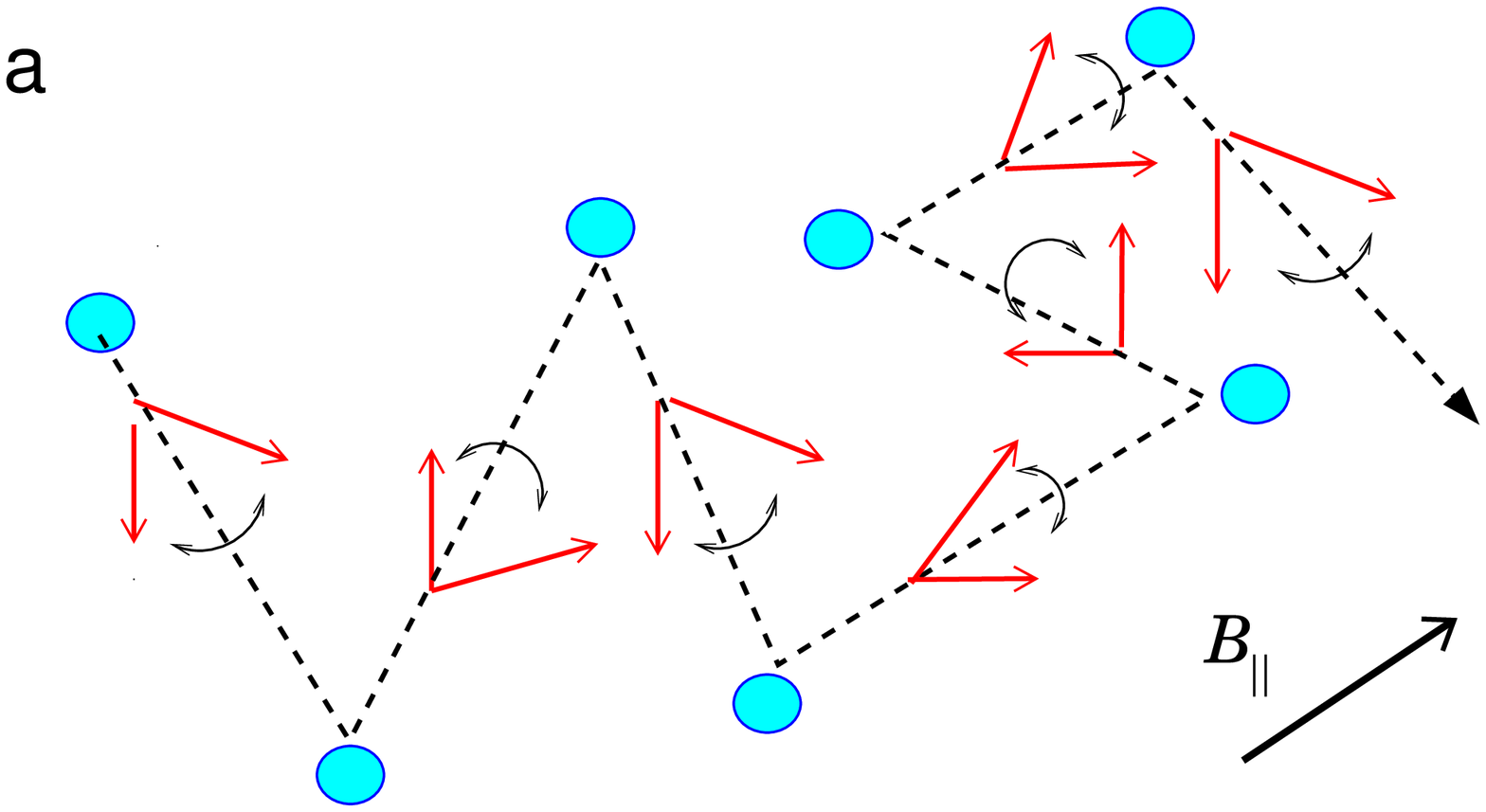} \\[0pt]
\includegraphics[width=8.0cm,angle=0]{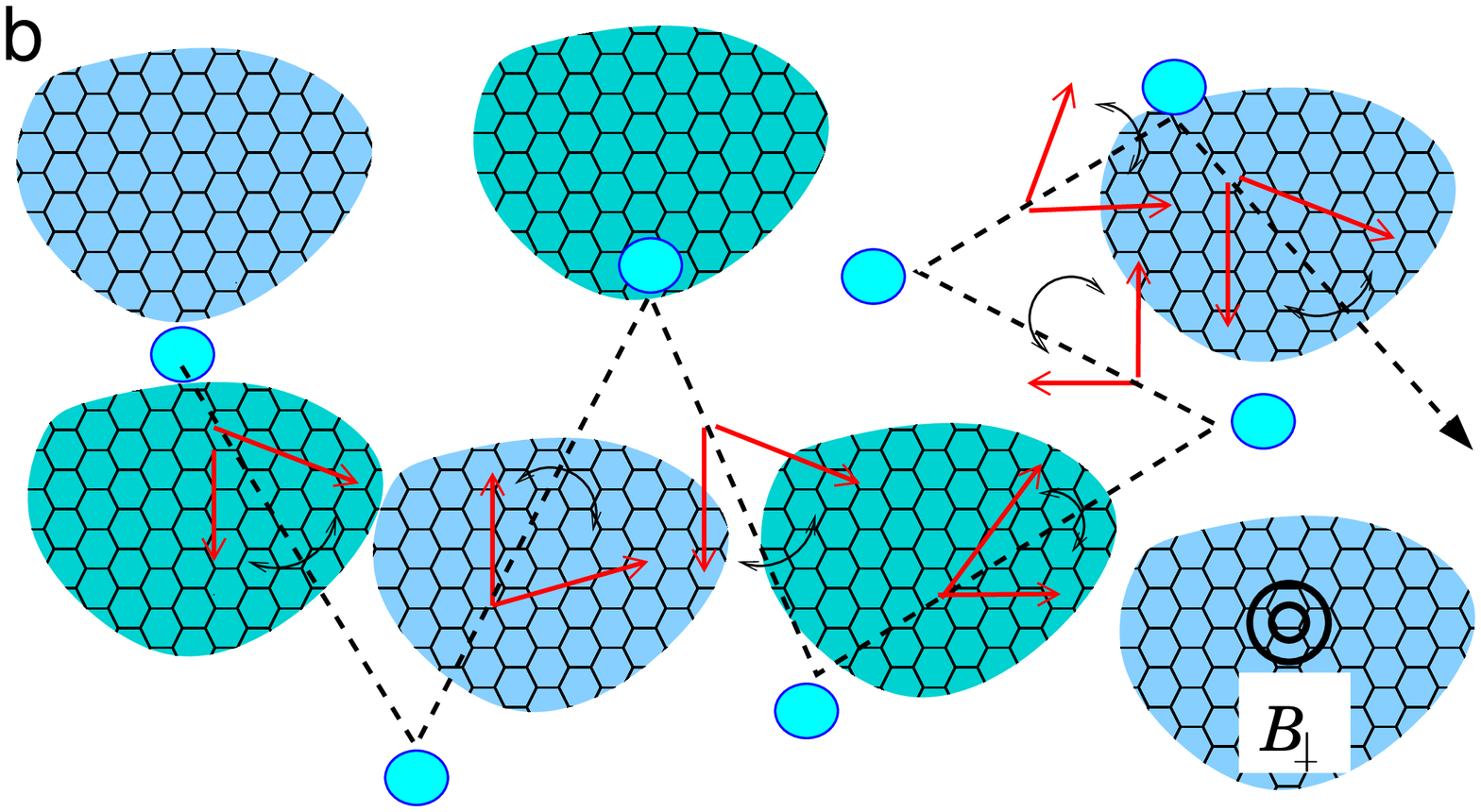}
\end{center}
\caption[fig]{(Color online).(a) Dyakonov-Perel spin relaxation: The SO
coupling induces a momentum dependent effective field $\vec{B}_{\parallel}$ which
changes direction randomly after scattering events, leading to spin
relaxation. $\vec{B}_{\parallel}$ is ``in plane'' so spins directed
perpendicular to the plane relax faster than as spins in the plane.
(b)Gauge-Field spin relaxation: Gauge fields due to
ripples (light and gray blue areas) induce an effective field $\mathcal{B}%
_{\perp}$, which for
a finite SO coupling leads to spin relaxation. $\vec{B}_{\perp}$ is
``out of plane'' so spins directed perpendicular to the plane relax slower than spins in the plane. }
\label{FIG1}
\end{figure}

Recently, spin transport and spin relaxation were studied in relatively
dirty graphene samples~\cite{HJPJW07,TTVJJvW08}. A spin relaxation length $%
\lambda_{\text{sf}}\sim 2\mu $m was measured at room temperature and it was
indicated that $\lambda_{\text{sf}}$ is proportional to the elastic mean free path
$l_{\text{el}}$, suggesting the EY mechanism to be dominant~\cite{HJPJW07,TTVJJvW08}%
. The measured spin relaxation length is weakly anisotropic, such that spins
\textquotedblleft out of plane\textquotedblright\ relax 20\% faster than
spins \textquotedblleft in plane\textquotedblright\ \cite{TTVJJvW08}. These
experiments motivate a study to see if known spin-relaxation mechanisms, or
possibly novel effects, dominate spin scattering in graphene.

In this Letter, we consider spin relaxation in intrinsic graphene
arising from three ingredients: (i) The SO interaction arises
from the band structure and can be enhanced by graphene
corrugations, (ii) spin isotropic scattering centers cause momentum
relaxation, (iii) topological lattice disorder induce gauge fields
that change the orbital motion. First, we study the Elliot-Yafet and
Dyakonov-Perel mechanisms for this model ignoring effects of gauge
fields. Second, we find that a unique interplay of SO and gauge
fields (GF) due to topological disorder causes spin relaxation.  Our main findings are that DP and GF
mechanisms are comparable and dominate the EY mechanism in intrinsic graphene.
Interestingly, the DP mechanism implies that spins out of plane
relax twice as fast as spins in plane, but GF exhibits the opposite
behavior, spins in plane relax faster than spins out of plane (see Fig. 1 for details). The
spin-relaxation anisotropy depends on the intricate competition
between the DP mechanism and GF. Our results are valid for relatively clean
graphene samples where \textit{e.g.} adatoms do not alter the
SO coupling.

\emph{Spin-orbit coupling.-} There are Rashba and Dresselhaus
\textquotedblleft like\textquotedblright\ SO interactions in
graphene and we disregard the latter \cite{KM05,HGB06,HMING06}. The total
Hamiltonian reads ($\hbar =1$) %
\begin{equation}
\mathcal{H}=\pm v_{\mathrm{F}}k\left( \vec{n}\cdot \hat{\mathbf{\sigma }}%
\right) -\frac{\Delta }{2}\left( \hat{\mathbf{\sigma }}\times \hat{\mathbf{s}%
}\right) _{z}\,,
\label{hamil}
\end{equation}%
where $=+(-)$ corresponds to the $K(K^{\prime })$ point, the spinor
basis for $K$ is $\Psi _{K}=\left( A_{\uparrow },A_{\downarrow
},B_{\uparrow },B_{\downarrow }\right) ^{T}$ and for $K^{\prime }$
the components are reversed, $\Psi _{K^{\prime }}=\left( B_{\uparrow
},B_{\downarrow },A_{\uparrow },A_{\downarrow }\right) ^{T}$, $A$
and $B$ denote the  lattice sites and $\uparrow$ and $\downarrow$
the electron spin. With this choice, the SO coupling is
identical for $K$ and $K^{\prime }$. $\vec{n}=(\cos
\theta ,\sin \theta )$, $\tan \theta =k_{y}/k_{x}$, $k=\sqrt{%
k_{x}^{2}+k_{y}^{2}}$ is quasi-particle momentum with respect to the $%
K(K^{\prime })$ corners of the hexagonal Brillouin zone, $\Delta =\Delta
_{\text{curv}}+\Delta _{\mathcal{E}}$ \cite{HGB06} is the dominant Rashba SO
coupling constant induced by curvature and/or an external electric fields
\cite{HGB06} and $\hat{\mathbf{\sigma }}=(\hat{\sigma}_{x},\hat{\sigma}_{y})$%
, $\hat{\mathbf{s}}=(\hat{s}_{x},\hat{s}_{y})$ correspond to Pauli
matrices in sublattice and spin space, respectively. The
eigenstates of (\ref{hamil}) for the $K$ valley are
\begin{equation*}
\left\vert \Psi _{k\xi s}\right\rangle ={N_{\xi s}}\left[ s\left(
\begin{array}{c}
c_{-s} \\
c_{s}e^{i\theta }%
\end{array}%
\right) i|\uparrow \rangle +\left(
\begin{array}{c}
c_{s}e^{i\theta } \\
c_{-s}e^{2i\theta }%
\end{array}%
\right) |\downarrow \rangle \right] e^{i\vec{k}\cdot \vec{r}}
\end{equation*}%
with energies $E_{k\xi s}=s\Delta/2+\xi D$, $D=\sqrt{(v_{%
\mathrm{F}}k)^{2}+\Delta ^{2}/4}$, where $s=\pm $ corresponds to the ${%
\uparrow (\downarrow )}$ spin states and $\xi =\pm $ denotes the
pseudospin degeneracy, $c_{k\xi s}=\sqrt{E_{k\xi s}}/\sqrt{2D}$,
$c_{s}\equiv c_{k\xi s} $ and $N_{\xi
s}=\sqrt{2 E_{k\xi s}}/(v_{\mathrm{F}}k)$. These
eigenstates are polarized in plane of the graphene layer. The spin
precession length is $l_{\text{prec}}=2\pi v_{F}/\Delta $.

\emph{Elliot-Yafet mechanism.-} The \textquotedblleft
Rashba\textquotedblright\ SO coupling can change the spin orientation during a
scattering event as is typical for the EY mechanism. We study this effect by
a decomposition into partial waves with a well defined orbital angular
momentum discussed in~\cite{phaseshift}. Neglecting mixing of the $K$ and $K^{\prime }$ valleys, an incoming wave
with total angular momentum, $\mathcal{L}\equiv \mathcal{I} i\partial
_{\theta }\pm \sigma _{z}/2+s_{z}/2$, is an eigenstate of Eq.~(\ref{hamil}):
\begin{align}
\Psi _{k\xi +}^{in}(r,\theta )& \equiv \left(
\begin{array}{c}
c_{-}J_{n}(kr)e^{in\theta } \\
c_{+}J_{n+1}(kr)e^{i(n+1)\theta }%
\end{array}%
\right) i|\uparrow \rangle  \notag \\
& +\left(
\begin{array}{c}
c_{+}J_{n+1}(kr)e^{i(n+1)\theta } \\
c_{-}J_{n+2}(kr)e^{i(n+2)\theta }%
\end{array}%
\right) |\downarrow \rangle  \label{Psi_EY}
\end{align}%
with $c_{\pm }$ as defined earlier and $J_{n}(x)$ is a Bessel function.
We analyze here weak scatterers, where the elastic scattering rate scales
as $\tau _{\text{el}}^{-1}\sim E_{F}$. Strong and resonant Coulomb scatterers \cite{phaseshift} induces a similar change in the spin orientation. We approximate the potential by a step function, $%
v(r)=V_{0}\left[ 1-\Theta (r-R)\right] $.  The wave function inside the potential well is a superposition
of two radial waves, finite at the origin and with different spin
orientations. In graphene, Rashba spin coupling entangles spin and pseudospin and the change of spin in a scattering event depends on the evolution of the pseudospin. Unlike conventional semiconductors\cite{ZFS04}, the spins in different angular momentum channels differ, complicating the definition of the amount of spin relaxation in a scattering event.
From the given incoming wave $\Psi ^{in}$ with spin parallel to the momentum
$\vec{k}$ and incident angle $\theta $, there are two possible outgoing
waves $\Psi _{k\xi +}^{out}(r,-\theta ),\Psi _{k\xi -}^{out}(r,\theta' )$,
which satisfy conservation of energy and momentum. These can be written in a
similar way as Eq.(\ref{Psi_EY}), where $\theta ^{\prime }\approx -\theta +\Delta \cot (\theta )/(v_{F}k)$. We then define
\begin{equation}
\mathcal{S}=\frac{\sum_{n}\left( r_{n}r_{0n}+r_{n}^{^{\prime
}}r_{0n}^{^{\prime }}\right) -\sum_{n}\left( r_{0n}^{2}+(r_{0n}^{^{\prime
}})^{2}\right) }{\sum_{n}\left( r_{0n}^{2}+(r_{0n}^{^{\prime }})^{2}\right) }%
\,,  \label{overlap}
\end{equation}%
where $r_{n}(r_{0n})$ and $r_{n}^{^{\prime }}(r_{0n}^{^{\prime }})$ are the
scattering amplitudes for a given angular momentum channel $n$ with
(without) SO coupling and for the two possible outgoing waves $\Psi
_{+}^{out},\Psi _{-}^{out}$ respectively. SO coupling changes the wavevector for one of the reflected waves, $%
k^{\prime }\approx k-\Delta /v_{F}$ . The leading reflection coefficient, $%
r_{n=0}$, in the absence of SO coupling depends on wavevector as $%
r_{0}(k)\sim V_{0}kR^{2}/v_{F}$\cite{phaseshift}, so that
$r_{n=0}(k^{\prime })-r_{n=0}(k)\sim V_{0}\Delta R^{2}/(v_{F})^{2}$
and $\mathcal{S}\sim \Delta /(v_{F}k)$. $\mathcal{S}$ vanishes if
the spin is conserved in the scattering event. If the changes
induced by a finite $\Delta$ are small, this quantity should be
proportional to the change in spin orientation during the scattering
process.  The change in spin orientation  at each collision is
$\Delta /(v_{F}k_{F})$.
The total change of the spin after $N_{coll}$ collisions is of order $\sqrt{%
N_{\text{coll}}}\Delta (v_{F}k_{F})^{-1}$. Dephasing occurs when $\sqrt{%
N_{\text{coll}}}\Delta (v_{F}k_{F})^{-1}=2 \pi$ after a time
$\tau_{\text{so}} =  \tau_\text{el} N_{\text{coll}}$,  where $\tau_\text{el}$
is the elastic scattering time, and the Elliot-Yafot spin relaxation
time is  $\tau _{\text{so}}^{\text{EY}}\sim (v_{F}k_{F})^{2}/\Delta
^{2}\times (l_{\text{el}}/v_{F})\sim (v_{F}k_{F})^{2}/\Delta
^{2}\times \tau_\text{el}$. The spin diffusion length is related to the spin
relaxation time by $\lambda_{\text{so}} = \sqrt{D
\tau_{\text{so}}}$, where $D=v_F^2 \tau/2$ so
$\lambda_{\text{so}}^{EY} \sim l_\text{el} (v_F k_F)/\Delta$.

\emph{The D'yakonov-Perel' mechanism.-} Between scattering events, the Rashba
SO coupling acts as an effective magnetic field in the plane, $\vec{B%
}_{\parallel }(\hat{\sigma})=\Delta (\hat{\sigma}\times \hat{e}_{z})/2$, on
the spins\cite{Dyakonov71,Dyakonov86}. The spin dynamics is a result of
spins precessing in a fluctuating in-plane magnetic field governed by the momentum. Elastic scattering randomizes the momentum and the associated
magnetic field. Averaging over many collisions, the spin orientation becomes
random after a time \cite{ZFS04}
\begin{equation}
\tau _{\text{so}}^{DP}\approx v_{F}l_{\text{el}}^{-1}/\Delta ^{2}\,,  \label{tau_DP}
\end{equation}%
\textit{e.g.} the spin-relaxation time is inversely proportional to
the elastic scattering time. The DP spin-diffusion length is
independent of the  mean free path $\lambda_\text{so}^{DP} = v_F /
(\sqrt{2} \Delta)=l_\text{prec}/(2 \sqrt{2} \pi)$. 

\emph{Effective gauge field.-} Topological lattice defects, strains, and
curvature change the hopping integrals between the lattice sites. These
effects are captured by an induced effective gauge field $\mathcal{A}(\vec{r}%
)$ which deflect electrons and change the electronic states at low energies 
\cite{NGPNG08,gauge}. $\mathcal{A}$ is related to the strain tensor $u_{ij}(%
\vec{r})$ describing topological lattice disorder , $\mathcal{A}_{x}\sim [u(%
\vec{r})_{xx}-u(\vec{r})_{yy}],\mathcal{A}_{y}\sim u(\vec{r})_{xy}$ and give
rise to a random out-of-plane magnetic field $\mathcal{B}_{\perp }=[\vec{%
\nabla}\times \mathcal{A}(\vec{r})]_{z}$ \cite{gauge}. We will demonstrate that
this out-of-plane orbital magnetic field together with the Rashba
SO coupling can induce an out-of-plane spin polarization. 

Let us first present a semiclassical argument for why the spin
polarization changes from in plane to out of plane as the gauge
field increases. $\mathcal{B}_{\perp }$ causes the electrons to move
in cyclotron orbits with a radius $r_c = 2 \epsilon/(e v
\mathcal{B}_{\perp })$, where $v$ is the velocity and $\epsilon$ is
the energy. A qualitative change in the electronic states occurs
when the cyclotron orbit is smaller than the spin precession length,
\textit{e.g.} when $2 \pi r_c = l_\text{prec}$. Using $v=v_F$, the
graphene dispersion $\epsilon = v_F k_F$, and introducing the
magnetic length $l_B=\sqrt{e/\mathcal{B}_{\perp }}$, there is a
transition when the magnetic length is shorter than
\begin{equation}
l_{B}^{c} = \sqrt{l_\text{prec} \lambda_F}/(2\sqrt{2} \pi) \, , \label{l_Btreshold}
\end{equation}
where $\lambda_F=2 \pi/k_F$ is the Fermi wavelength. In the low field
regime, $l_B \gg l_{B}^{c}$, in-plane momentum is a good quantum
number and via the SO interactions spins are polarized
in plane. In contrast, in the strong field regime, $l_B \ll
l_{B}^{c}$, out-of-plane angular momentum is a good quantum number,
and the SO coupling changes the polarization of the states
to out of plane.

We will now carry out a quantum mechanical calculation which  will
give more details and confirm our arguments above with the threshold
value Eq. (\ref{l_Btreshold}) for the gauge field. Inserting the gauge
fields in the Hamiltonian in Eq. (\ref{hamil}), results in a
$4\times 4$ Hamiltonian for one valley $K$,
\begin{equation}
\mathcal{H}=\left(
\begin{array}{cccc}
0 & 0 & v_{\mathrm{F}}\hat{\Pi} & 0 \\
0 & 0 & -i\Delta & v_{\mathrm{F}}\hat{\Pi} \\
v_{\mathrm{F}}\hat{\Pi}^{\dag } & i\Delta & 0 & 0\label{H_GF} \\
0 & v_{\mathrm{F}}\hat{\Pi}^{\dag } & 0 & 0%
\end{array}%
\right)  \, ,
\end{equation}%
where $\Pi =P_{x}-iP_{y}$, $P=-i\nabla +e\mathcal{A}$ and a similar
Hamiltonian can be written for $K^{\prime }$, where the sign of the gauge
field is reversed. First, we consider the solution of
the Dirac equation when $\Delta=0$. The results are
gauge-invariant and we choose the Landau gauge $\mathcal{A}_{x}=0,\mathcal{A}%
_{y}=\mathcal{B}_{\perp }x$ and we consider a homogenous magnetic field to illustrate the main effect. The wave functions are
\begin{equation}
\Phi _{n,\uparrow ,\downarrow }=\left(
\begin{array}{c}
\mp i\phi _{n-1}(x-x_{0}) \\
\phi _{n}(x-x_{0})%
\end{array}%
\right) e^{iky}|\uparrow ,\downarrow \rangle ,
\end{equation}%
with eigenenergy $\epsilon _{n}=\pm v_{\mathrm{F}}\sqrt{2|n|}/l_{B}$
in terms of solutions of a particle in an harmonic oscillator
potential $\phi _{n}(x-x_{0})$. The two components of $\Phi
_{n,\uparrow ,\downarrow }$ correspond to the amplitudes in the two
graphene sublattices and $x_{0}=kl_{B}^{2}$ is the Landau level
guiding center.

The Rashba SO coupling induces an interaction between
electrons with spin up in one sublattice and electrons with spin
down in the other sublattice. Therefore it is convenient to express
the Hamiltonian (\ref{H_GF}) in the basis $\phi _{n-1}\left[
1,0\right]
^{T}|\uparrow \rangle $, $\phi _{n}\left[ 0,1\right] ^{T}|\uparrow \rangle $%
, $\phi _{n}\left[ 1,0\right] ^{T}|\downarrow \rangle $, $\phi _{n+1}\left[
0,1\right] ^{T}|\downarrow \rangle $, $\mathcal{H=\tilde{H}}v_{\mathrm{F}}%
\sqrt{2}/l_{B}$:
\begin{equation}
\mathcal{\tilde{H}}=\left(
\begin{array}{cccc}
0 & i\sqrt{|n|} & 0 & 0 \\
-i\sqrt{|n|} & 0 & -i\tilde{\Delta} & 0 \\
0 & i\tilde{\Delta} & 0 & i\sqrt{2|n+1|} \\
0 & 0 & -i\sqrt{|n+1|} & 0%
\end{array}%
\right) ,
\end{equation}%
where $\tilde{\Delta}=\Delta l_{B}/(v_{F}\sqrt{2})$. The eigenenergies of $%
\mathcal{\tilde{H}}$ are:

\begin{equation*}
\tilde{\epsilon}_{n}^{2}=\sigma [ 1+\tilde{\Delta}^{2}+2n+s\sqrt{\left(
1+\tilde{\Delta}^{2}\right) ^{2}+4n\tilde{\Delta}^{2}}] /2,
\end{equation*}%
where, $\sigma =+$ ($\sigma =-$) denotes electron (hole) like excitations
and $s=+$ ($s=-$) denotes spin. Let us consider the expectation value of the
out-of-plane spin polarization of these states, $p_{z}$. For $n=0$ there are three
physical states of which one has polarization $(\hbar /2)(1-\tilde{\Delta}%
^{2})(1+\tilde{\Delta}^{2})$ and two have polarizations $-(\hbar /2)/(1+%
\tilde{\Delta}^{2})$. For all states where $n\geq 1$ the polarizations are
\begin{equation}
p_{z}=s\frac{\hbar }{2}[\left( 1+\tilde{\Delta}^{2}\right) ^{2}+4n%
\tilde{\Delta}^{2}] ^{-1/2}
\label{Pz}
\end{equation}%
The spin polarization differs between the lowest and highest Landau levels and
the transition roughly occurs when $4\sqrt{n}\tilde{\Delta} \approx 1$%
. Transport is governed by states at the Fermi energy
$\epsilon_n \approx v_F \sqrt{2|n|}/l_B \rightarrow v_F k_F$ and the condition $4%
\sqrt{n}\tilde{\Delta}\approx 1$ can be rewritten in terms of a critical
value for the magnetic length, which \textit{exactly} agrees with our
semiclassical estimate for $l_B^{c}$ in Eq. (\ref{l_Btreshold}).
Equation (\ref{Pz}) demonstrates that when $l_{B}\ll l_{B}^{(c)}$ the states are fully
out-of-plane polarized, but the out-of-plane polarization vanishes
when the gauge field is weak $l_{B}\gg l_{B}^{(c)}$. The energy splitting is of the order $\Delta $ for
weak gauge fields and reduced by a factor $\left[
l_{B}^{(c)}/l_{B}\right] ^{2}$ for stronger gauge fields. \ \qquad

This change in the polarization direction of the eigenstates with increasing
gauge fields has consequences for the spin relaxation. When the gauge fields
vanish, the effective magnetic field is in plane so that spins out of plane
relax twice as fast as spins in plane. In the regime around $%
l_{B}=l_{B}^{(c)}$, the in-plane and out-of-plane components of the
effective magnetic field are comparable and we expect the spin-relaxation
anisotropy to be reduced and eventually exhibit the opposite behavior, spins
in plane relax faster than spins out of plane.

\emph{Experimental consequences.-}  We use typical
parameters for graphene, $\hbar v_F = 5.3\times 10^{-10} eV m$ and the
enhanced  Rashba coupling for a ripple of radius $R=100 nm$ is
$\Delta_{R=100nm}  =  1.7 \times 10^{-5} eV$ \cite{HGB06}. The Fermi wavelength
depends on the electron doping $n$, $\lambda _{F}=\sqrt{2\pi /n}$ .
In Ref. \cite{HJPJW07} $n \approx 3.6 \times 10^{16}$m$^{-2}$ so
$\lambda _{F}\approx 13 $nm and $l_{\text{el}}=36$ nm, but
considerably larger mean free paths have been measured and should be
expected in clean systems in the future.

Comparing the DP and EY relaxation mechanisms,
$\tau_{\text{so}}^{EY}\approx
(v_{F}k_{F})^{2}/\Delta_{\text{so}}^{2}\tau _{p}$ and
$\tau_{\text{so}}^{DP}\approx \tau _{p}^{-1}/\Delta
_{\text{so}}^{2}$, we find $\tau_{\text{so}}^{EY}/
\tau_{\text{so}}^{DP} \approx \tau_{p}^{2} (v_{F} k_{F})^{2} \sim
(k_{F} l_{\text{el}} )^{2}$. Typically $(k_F l_\text{el})^2 \gg1$,
\textit{e.g} in Ref. \cite{HJPJW07} $(k_F l_\text{el})^2 \approx 300$,
so the DP mechanism is much more important that the EY mechanism.

The experimental trend that the spin-relaxation length is
proportional to the mean free path~\cite{HJPJW07,TTVJJvW08} is encouraging since it suggests clean systems should
have a very long spin-relaxation length. However, this is at odds
with our results for intrinsic graphene where the spin-relaxation
length only weakly depends on the mean free path. The good news is
that the computed spin relaxation length is long, we find $l_{prec}
= 2 \pi\hbar v_F/\Delta_{R=100nm}   =  190 \mu m$ and hence the DP
spin relaxation length is $\lambda_{\text{sf}}^{DP}=
l_{\text{prec}}/(2 \pi \sqrt{2}) \sim 20 \mu$m. On the other hand,
in Ref. \cite{HJPJW07} $\lambda _{sf}\sim 1.3-2 \mu$m. Our theory neither quantitively nor
qualitatively explains the experiments in Ref. \cite{HJPJW07}. Adatoms such as2 hydrogen that locally enhance the SO
interaction could be responsible for the discrepancy since the mobility
of the experimental samples is relatively low\cite{CastroNeto:adatom}.

Our theory applies to cleaner, intrinsic graphene, possibly with
less  adatoms, where we predict novel spin-relaxation anisotropy
effects. We expect it is possible to obtain longer
spin-relaxation lengths than in Ref. \cite{HJPJW07,TTVJJvW08}, but
that one eventually will enter the intrinsic regime where the DP
mechanism prevents a further enhancement.
Furthermore, in intrinsic graphene, gauge fields due to ripples are
important. From the parameters above, we find a threshold magnetic
length of $l_{B}^{(c)}\approx 200$ nm.  Surface corrugations give
rise to effective magnetic lengths of the order $l_{B}\approx 100$nm \cite{gauge} so already at these electron densities gauge fields are
important and reduce the spin relaxation with respect to DP
mechanism.  Lower electron densities, which are feasible, should
enhance the effects of the gauge fields. In ultra clean systems, $l_\text{el} > l_B$, we expect $\tau_\text{so}^{DP} \rightarrow v_F l_B^{-1}/\Delta^2$ and the spin-relaxation time saturates. We encourage experiments in clean graphene, which will further elucidate the interplay between
SO coupling, momentum relaxation, and gauge fields.

We acknowledge discussions with M. Popinciuc and
B.J. van Wees. D. H.-H. and A. B. acknowledge support by the Research Council of Norway, Grants Nos. 158518/143 and 158547/431. F.G. acknowledges support from
MEC (Spain) through grant FIS2005-05478-C02-01 and CONSOLIDER CSD2007-00010,
by the Comunidad de Madrid, through CITECNOMIK, CM2006-S-0505-ESP-0337.


\end{document}